\documentclass[preprint]{aastex}

\slugcomment{Submitted to APJL}

\shorttitle{TWO FAST ACCELERATION PHASES}

\shortauthors{Song et al.}

\begin{document}
\title{ACCELERATION PHASES OF A SOLAR FILAMENT DURING ITS ERUPTION}
\author{H.Q. SONG\altaffilmark{1}, Y. CHEN\altaffilmark{1}, J. ZHANG\altaffilmark{2}, X. CHENG\altaffilmark{3}, H. Fu\altaffilmark{1}, AND G. LI\altaffilmark{4}}

\affil{1 Shandong Provincial Key Laboratory of Optical Astronomy
and Solar-Terrestrial Environment, and Institute of Space
Sciences, Shandong University, Weihai, Shandong 264209, China}
\email{hqsong@sdu.edu.cn}

\affil{2 School of Physics, Astronomy and Computational Sciences,
George Mason University, Fairfax, VA 22030, USA}

\affil{3 School of Astronomy and Space Science, Nanjing
University, Nanjing, Jiangsu 210093, China}

\affil{4 Department of Space Science and CSPAR, University of
Alabama in Huntsville, Huntsville, AL 35899, USA}


\begin{abstract}
Filament eruptions often lead to coronal mass ejections (CMEs),
which can affect critical technological systems in space and on
the ground when they interact with the geo-magnetosphere in high
speeds. Therefore, it is an important issue to investigate the
acceleration mechanisms of CMEs in solar/space physics. Based on
observations and simulations, the resistive magnetic reconnection
and the ideal instability of magnetic flux rope have been proposed
to accelerate CMEs. \textbf{However, it remains elusive whether
both of them play a comparable role during a particular eruption.}
It has been extremely difficult to separate their contributions as
they often work in a close time sequence during one fast
acceleration phase. Here we report an intriguing filament eruption
event, which shows two apparently separated fast acceleration
phases and provides us an excellent opportunity to address the
issue. Through analyzing the correlations between velocity
(acceleration) and soft (hard) X-ray profiles, we suggest that the
instability and magnetic reconnection make a major contribution
during the first and second fast acceleration phases,
respectively. Further, we find that both processes have a
comparable contribution to accelerate the filament in this event.

\end{abstract}


\section{INTRODUCTION}
Coronal mass ejections (CMEs) are the most energetic eruptions in
the solar system and can affect critical technological systems in
space and on the ground when they interact with the
geo-magnetosphere in high speeds ranging from several hundred to
even more than one thousand kilometers per second. Therefore, it
is an important issue to investigate the acceleration mechanisms
of CMEs in solar/space physics. It is generally accepted that CMEs
are driven by magnetic flux rope (MFR) eruptions (Chen 2011).
However, we can not observe MFR structures directly in the corona
because no instrument can provide the high quality measurement of
the coronal magnetic filed at present. Several lines of
observations in the lower corona have been proposed as the proxies
of MFRs, e.g., filaments/prominences (Rust \& Kumar 1994), coronal
cavities (Wang \& Stenborg 2010), sigmoids (Titov \& D{\'e}moulin
1999; McKenzie \& Canfield 2008), and hot channels (Zhang et al.
2012; Song et al. 2014a, 2014b, 2015). Resolving the dynamics of
these structures are critical to our understanding of the CME
acceleration process.

\textbf{There exist two important magnetic energy release
mechanisms: one is the resistive magnetic reconnection process
(Carmichael 1964; Sturrock 1966; Hirayama 1974; Kopp \& Pneuman
1976) and the other is the ideal global magnetohydrodynamic (MHD)
MFR instability (van Tend \& Kuperus 1978; Priest \& Forbes 1990;
Forbes \& Isenberg 1991; Isenberg et al. 1993; Forbes \& Priest
1995; Hu et al. 2003; Kliem \& T\" or\" ok 2006; Fan \& Gibson
2007; Chen et al. 2007a, 2007b; Olmedo \& Zhang 2010). Both
mechanisms are supported by observations. For example, good
correlations exist between CME speed (acceleration) and the soft
X-ray (hard X-ray and microwave) profiles of associated flares
(Zhang et al. 2001; Qiu et al. 2004; Mari\v ci\'c et al. 2007). In
addition, studies showed that the extrapolated magnetic flux in
the flaring region was comparable with the magnetic flux of the
MFR reconstructed from in situ data (Qiu et al. 2007). All these
CME-flare association studies support that reconnections play an
important role in accelerating CMEs. On the other hand,
statistical studies showed that the projected speed in the sky
plane and kinetic energy of CMEs only had weak correlations with
the peak values of their associated X-ray flares (Yashiro et al.
2002; Vr\v snak et al. 2005). These observations support that the
ideal MHD instability also makes significant contributions to the
CME acceleration besides the magnetic reconnection.}

\textbf{The theoretical and simulation studies also support that
the instability and reconnection can accelerate the CMEs
(D{\'e}moulin \& Aulanier 2010; Roussev et al. 2012; Amari et al.
2014). Generally, the instability plays an important role in
triggering and accelerating the MFR, and then the magnetic
reconnection accelerates the MFR further and allows the process to
develop continuously (Priest \& Forbes 2002; Lin et al. 2003).}

\textbf{However, it remains elusive whether both mechanisms have a
comparable contribution to the acceleration in a particular event
as they usually accelerate CMEs in a close time sequence. In this
letter, we address this issue through analyzing a filament
eruption with two apparently separated fast acceleration phases,
instead of one as usual.} The relevant observations and results
are described in Section 2. Section 3 presents the related
discussion, which is followed by a summary in Section 4.

\section{OBSERVATIONS AND RESULTS}

\subsection{Instruments}
The eruption process was recorded by the Atmospheric Imaging
Assembly (AIA) telescope (Lemen et al. 2012) on board the
\textit{Solar Dynamic Observatory (SDO)}. AIA has 10 narrow UV and
EUV passbands with a cadence of 12 s, a spatial resolution of
1.2$''$, and an FOV of 1.3 $R_\odot$. The AIA images shown in this
letter (Figure 1 and Supplementary Movie) are a small portion of
the original full size images.

The soft and hard X-ray (SXR and HXR) data shown in Figure 3 are
from \textit{Geostationary Operational Environment Satellite
(GOES)} and the \textit{Reuven Ramaty High Energy Solar
Spectroscopic Imager (RHESSI:} Lin et al. 2002), respectively.
\textit{GOES} provides the integrated full-disk SXR emission from
the Sun, which are used to characterize the magnitude, onset time,
and peak time of solar flares. \textit{RHESSI} provides the HXR
spectrum and imaging of solar flares.

\subsection{Overview of the filament eruption}
The filament eruption originated from NOAA Active Region 12151
located at the heliographic coordinates $\sim$S09E76 on 2014
August 24. This eruption produced an M5.9 class SXR flare on
\textit{GOES} scale, which started at 12:00 UT and peaked at 12:17
UT. A CME (linear velocity 417 km s$^{-1}$) associated with it was
recorded by the Large Angle Spectroscopic Coronagraph (LASCO,
Brueckner et al. 1995) on board the \textit{Solar Heliospheric
Observatory}. We inspect the AIA images and find that the filament
appeared in all bandpasses corresponding to the coronal and
chromospheric temperatures, which indicate that it has a
multi-thermal nature. The eruption process snapshots observed with
304 \AA~($\sim$0.05 MK), 171 \AA~($\sim$0.6 MK) and 335
\AA~($\sim$2.5 MK) are presented at the top, middle and bottom
panels of Figure 1, respectively. The time when the image was
taken is shown at the top of each panel. The arrows depict the EUV
brightening positions, where the plasma was heated by magnetic
reconnection. The full AIA image sequences of the eruption in six
passbands are provided in the Supplementary Movie.

In order to clearly display the rising motion of the filament, a
slice-time plot was constructed with base-difference images of
304~\AA\ along the dotted line in panel b of Figure 1, as
presented in Figure 2. Around 12:16:00 UT, the filament apex moved
out of the field of view (FOV) of AIA, so no further kinematic
evolution analysis was possible after that time. Through the
time-stacking plot, we measure the filament height with time for a
careful kinematic analysis as described in next subsection.

\subsection{Acceleration processes of the filament}
The kinematic information of the filament is obtained by analyzing
AIA 304~\AA~ base-difference images. We carefully inspect the
images and identify the filament leading edges along the slice in
Figure 1(b). The heights are measured from the projected distance
of the leading edges from the flare location. The uncertainty of
the height measurement is about 4 pixels (2 Mm), which are
propagated to estimate the velocity errors in the standard way.
Based on the height-time measurements, the velocities are derived
from a numerical derivative method that employs Lagrangian
interpolation of three neighboring points, a piecewise approach to
calculate the velocity (Zhang et al. 2001; Cheng et al. 2013; Song
et al. 2014a, 2014b). Then the filament acceleration is calculated
with the same method based on the calculated velocity profile.

Figure 3 presents the kinematic analysis results. The entire
eruption process as seen by AIA can be divided into three distinct
phases as shown in panel a: an initial slow-rise phase, and then
two fast acceleration phases. The vertical green solid lines on
the left and right demarcate the start time of the first and
second fast acceleration phases, respectively. In addition, the
SXR flux with time can also be divided into a slow-rise phase and
two impulsive phases. The vertical black dotted line denotes the
onset of the first impulsive phase of X-ray flare. Apparently, the
right green solid line also marks the onset of the second
impulsive phase of the flare.

The slow-rise phase of the filament lasted for about 7 min from
$\sim$12:00 to $\sim$12:07 UT; at the end of this phase, the
velocity increased to $\sim$40 km s$^{-1}$ with an averaged
acceleration of 95 m s$^{-2}$. The first fast acceleration phase
started around 12:07:19 UT and ended around 12:11:31 UT when the
velocity reached to $\sim$450 km s$^{-1}$ with an averaged
acceleration of $\sim$1626 m s$^{-2}$. Then the velocity decreased
slightly during the following $\sim$1.5 minutes. The second fast
acceleration phase started around 12:13:00 UT. As the filament
moved out of the AIA FOV, we can trace a part of the second
acceleration phase. The final velocity we deduce is $\sim$700 km
s$^{-1}$ around 12:15:40 UT. Therefore, the averaged acceleration
for this phase is $\sim$1600 m s$^{-2}$. The horizontal red solid
lines in Figure 3 depict these two fast acceleration periods. As
mentioned, the SXR flux with time has two impulsive phases, which
indicate there exist two obvious reconnection processes. This
point is further confirmed by the derivation of the SXR flux with
two peaks as shown in panel b. The two red dots in panels b and c
depict the positions of two peaks, which show that the first
obvious reconnection is weak compared with the second one.

The onset of CME acceleration phase often coincides well with the
onset of accompanying flares (Zhang et al. 2001). For our event,
it is obvious that the velocity and SXR flux profiles are tightly
consistent during their slow-rise phase and second fast
acceleration phase. However, the onset time of the first fast
acceleration phase of filament (12:07:19 UT) is obviously ahead of
the first impulsive phase of flare (12:09:19 UT) for two minutes,
which means the filament was accelerated before the first obvious
reconnection onset. Thanks to AIA's high cadence, such short
difference is otherwise difficult to observe and measure.

Further, we plot the HXR flux versus time in panel c, along with
the filament acceleration profile. The error bars of acceleration
are large and not shown here. Note the HXR rates fell after 12:05
UT when the thin attenuators moved into the detector FOV, and it
fell suddenly again before 12:18 UT when \textit{RHESSI} went into
night. We made two Gaussian fitting for the HXR flux as presented
with two thin blue solid curves in the panel. We believe the two
fitted Gaussian shapes correspond to the two obvious reconnection
processes according to the derivation of SXR flux in panel b. The
observed HXR peaks in the middle of the two fitted peaks should be
due to the overlapping effect of HXR emission from these two
reconnections, which is supported by the accumulation of the two
fitted curves as shown with the thick blue solid curve. Panel c
shows that the first fast acceleration profile did not have any
correlations with the HXR flux, even the acceleration continued to
decrease after the reconnection onset. They tightly correlated
with each other only during the second fast acceleration process.
\textbf{Therefore, we suggest that the first fast acceleration
process was induced by the instability (Cheng et al. 2013), while
the second one was accelerated mainly through the subsequent
magnetic reconnection (Priest \& Forbes 2002; Lin et al. 2003).}
The AIA 335~\AA~observations present a kink morphology during the
eruption (See Figure 1(h)), indicating that the filament might be
accelerated through the MFR kink instability (Cheng et al. 2014).
Our study suggests that both mechanisms have a comparable
contribution to the CME acceleration, which is consistent with the
MHD simulation results (Chen et al. 2007b).

\section{DISCUSSION}
Our observational results have significant theoretical
implications. \textbf{The event provides us an excellent
opportunity to compare the contributions of different acceleration
mechanisms.} During the eruption, the filament first exhibits a
slow-rise phase, followed by two fast acceleration phases. For the
slow-rise phase, the acceleration might be induced by the initial
weak quasi-separatrix-layer (QSL) reconnection (Cheng et al.
2013), i.e., reconnections at the interface between filament and
its surrounding corona. The brightening as shown with arrows in
Figures 1(a) and (d) indicated that there existed reconnection
during that period (Cheng et al. 2013). Another possibility is
that the slow-rise phase was due to the initial acceleration stage
caused by the MFR kink instability. It is not possible to
distinguish their contributions at this phase.

However, for the following two fast acceleration phases, we
suggest that they are attributed to the instability and the
reconnection, respectively. It can be seen that the first obvious
reconnection didn't result in any considerable filament
acceleration. This might be explained by two possible factors:
first, the reconnection is weaker compared to the second one
according to their HXR peak values as mentioned above.
\textbf{This weak reconnection might be insufficient to accelerate
the MFR as it can not weaken the tension force of the overlying
magnetic loops fast enough, and may even lead to the MFR
deceleration as calculated by Lin \& Forbes (2000) and Lin
(2002)}; second, it is likely that this reconnection is still the
QSL reconnection, as told from the HXR data of \textit{RHESSI}.
The locations of the HXR sources are shown as black contours at
70\%, 80\%, and 90\% of the maximum in the 25-50 keV band (Figures
1(b),(c),(e),(f)). The sources are mainly at the two footpoints of
the filament during the first obvious reconnection (Figures
1(b),(e)), which is consistent with the QSL reconnection. The QSL
reconnection can heat the filament (as depicted with arrows in
Figures 1(b) and (e)) (Cheng et al. 2013) and produce high energy
electrons that escape along two legs of the filament and produce
HXR emission at the footpoints, but it does not contribute to
accelerate filament. On the other hand, the HXR sources locate
mainly between the two footpoints during the second obvious
reconnection (Figures 1(c),(f)), which indicates the reconnection
mainly took place in the current sheet connecting the filament to
the flare loop. This reconnection can accelerate filament/MFR
(Carmichael 1964; Sturrock 1966; Hirayama 1974; Kopp \& Pneuman
1976; Lin \& Forbes 2000; Lin 2002; Chen 2011) and produce
energetic electrons that hit the flare loop to produce the
\textit{RHESSI} HXR source.

If there was not the first obvious magnetic reconnection, and the
second obvious reconnection took place quickly after the kink
instability onset, we will not be able to distinguish the
contributions from the instability and reconnection, and we should
observe only one fast acceleration phase like most events.
Therefore, we point out that the time difference between the
instability onset and the subsequent reconnection in the current
sheet should be long enough to distinguish their contributions. We
conjecture that this time difference is very short in most events,
which might be one important reason that similar events with two
fast acceleration phases are rare.

\section{SUMMARY}
\textbf{A filament eruption associated with an M5.9 class flare
was well observed by the AIA at the southeast limb of the Sun on
2014 August 24, which presented two fast acceleration phases
during the eruption and provided us a perfect opportunity to
compare the contributions of different acceleration mechanisms of
CMEs/MFRs during a particular event.} Based on the detailed
analysis of the relations between velocity (acceleration) and SXR
(HXR) profiles, we suggest that the instability and magnetic
reconnection make a major contribution during the first and second
fast acceleration phases, respectively. The averaged acceleration
for the first phase is $\sim$1626 m s$^{-2}$, similar to that of
the second phase ($\sim$1600 m s$^{-2}$). Therefore, both
instability and reconnection play a comparable role to accelerate
the filament in this event, which is consistent with the MHD
simulation results (Chen et al. 2007b). We summarize the two fast
acceleration phases as follow: the instability takes place first
in a catastrophic manner, then the MFR is accelerated by the
Lorentz force (Chen et al. 2007b; Amari et al. 2014). The magnetic
energy is mainly transformed into the kinetic energy of MFR. In
the meanwhile, current sheet is bound to form as the eruptive MFR
drags magnetic field lines outwards, which provides proper site
for fast magnetic reconnection. The subsequent reconnection
produces a further acceleration of the MFR, i.e., the second fast
acceleration phase.

\acknowledgments We thank the referee for his/her valuable
comments that improved the manuscript. SDO is a mission of NASA's
Living With a Star Program. This work is supported by the 973
program 2012CB825601, NNSFC grants 41274177, 41274175, and
41331068. J.Z. is supported by US NSF AGS-1249270 and NSF
AGS-1156120.

\clearpage

\begin{figure}
\epsscale{1.0} \plotone{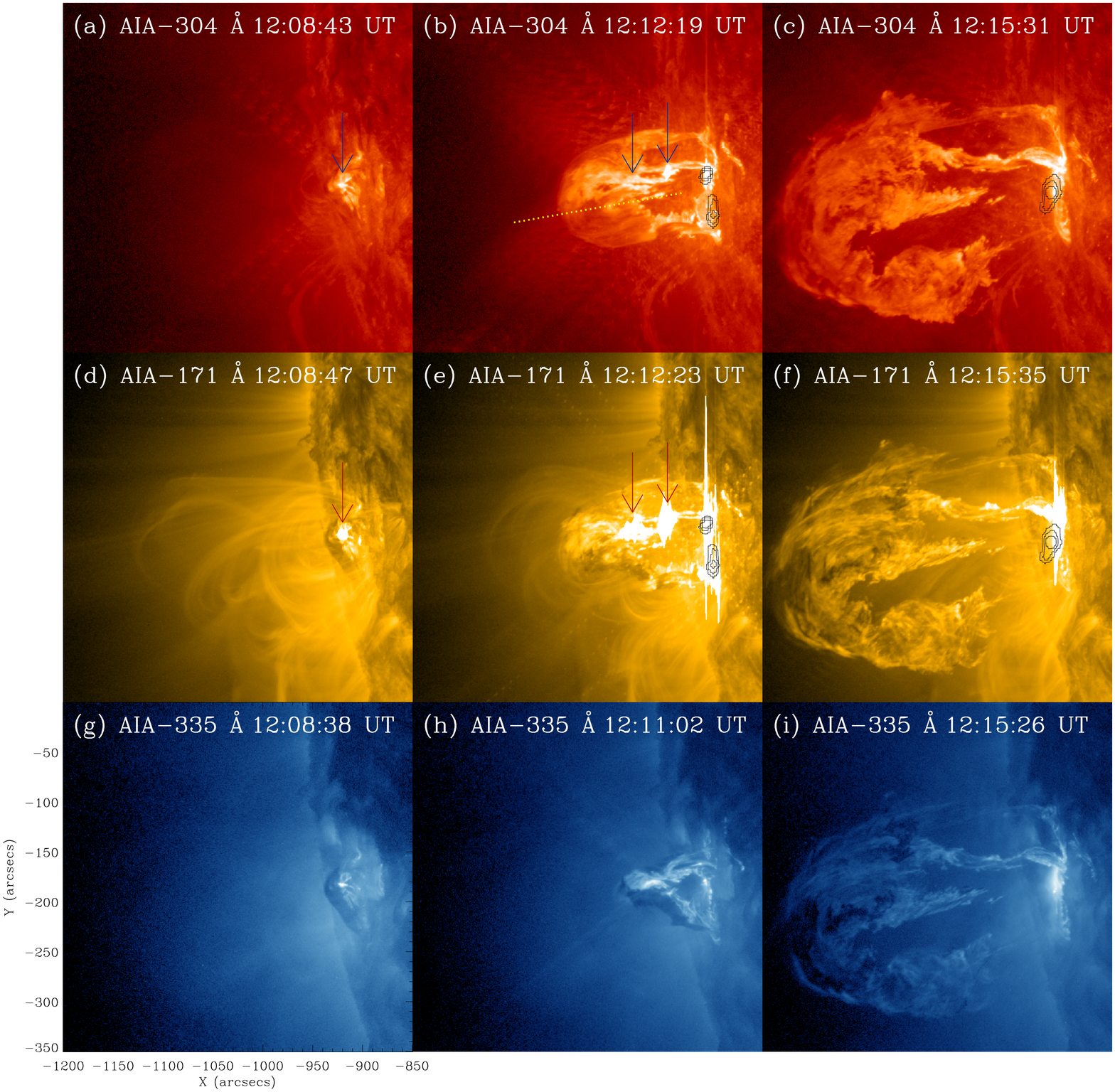} \caption{Filament eruption event
on 2014 August 24 observed in AIA images. (A color version and
animation of this figure are available in the online
journal.)\label{Figure 1}}
\end{figure}

\begin{figure}
\epsscale{1.0} \plotone{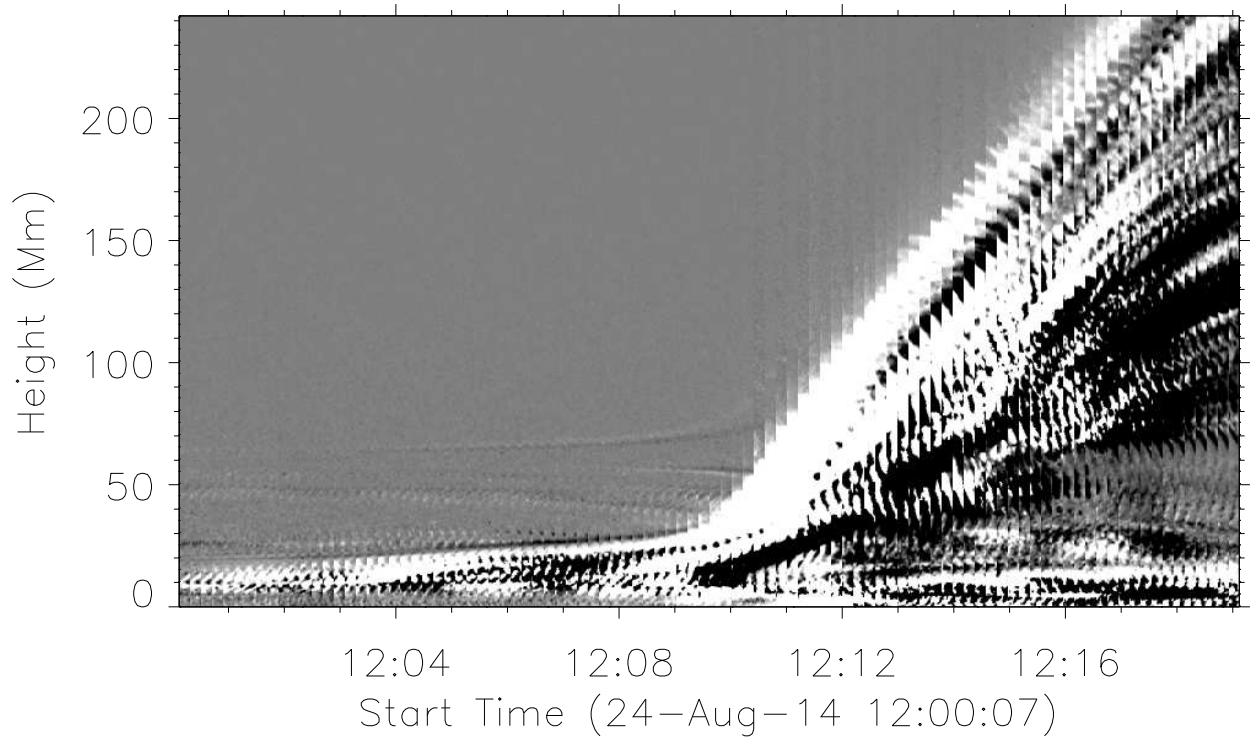} \caption{The constructed
slice-time plots illustrating the rising motion of the filament.
\label{Figure 2}}
\end{figure}

\begin{figure}
\epsscale{0.7} \plotone{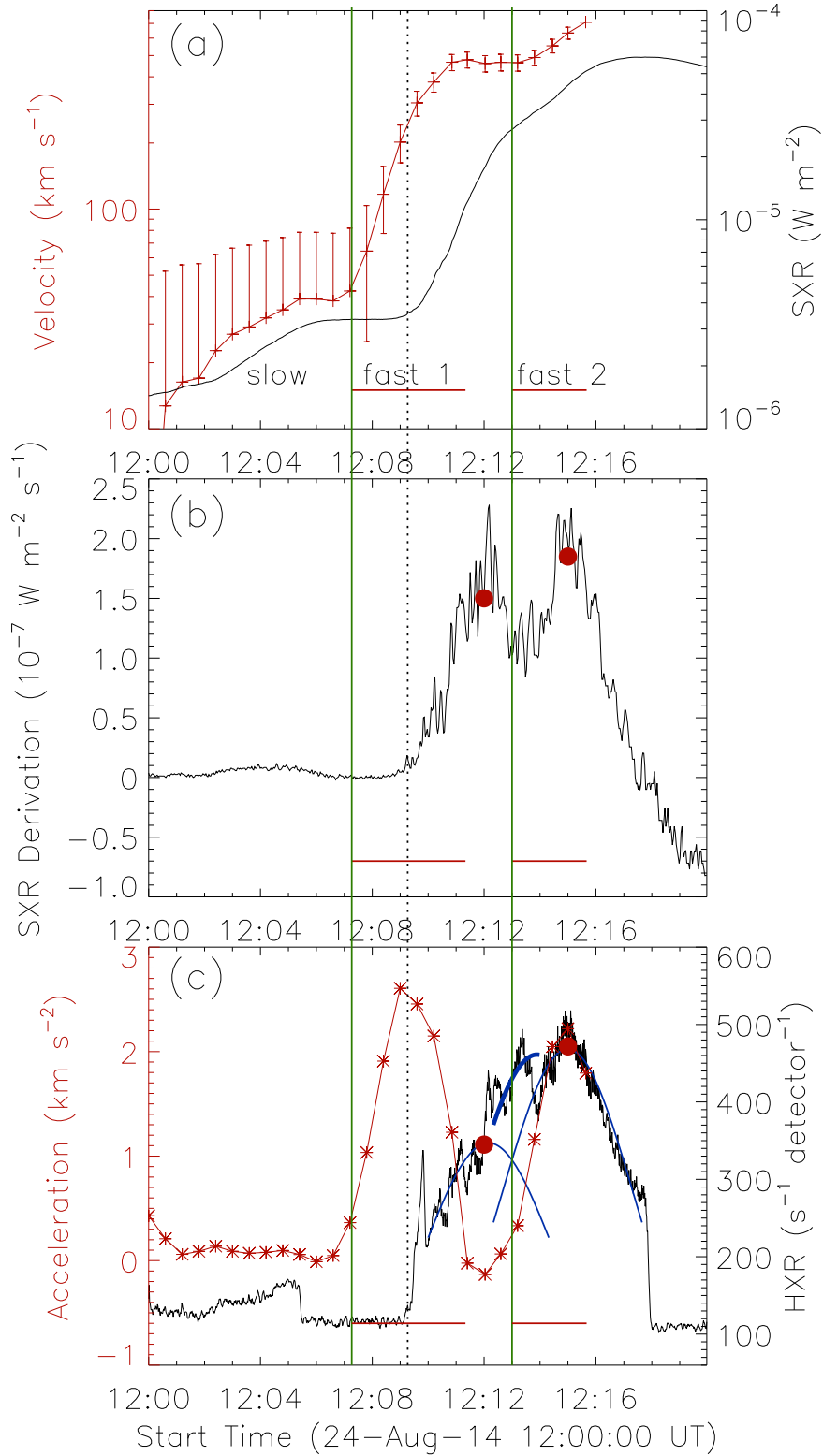} \caption{(a) The velocity-time
plots of the filament (red solid line), along with \textit{GOES}
SXR 1--8~{\AA}\ flux profiles of the accompanying flare. (b) The
derivation of the SXR flux. (c) The acceleration-time plots (red
solid curve), along with \textit{RHESSI} HXR 12-25 keV (black
solid curve) and their gaussian fittings (blue solid curve). (A
color version of this figure is available in the online journal.)
\label{Figure 3}}
\end{figure}


\begin{thebibliography}{}
\bibitem[Amari et al.(2014)]{2014Natur.514..465A}
Amari, T., Canou, A.,\& Aly, J.-J.\ 2014, \nat, 514, 465


\bibitem[Brueckner et al.(1995)]{1995SoPh..162..357B}
Brueckner, G.~E., Howard, R.~A., Koomen, M.~J., et al.\ 1995,
\solphys, 162, 357




\bibitem[Carmichael(1964)]{1964NASSP..50..451C}
Carmichael, H.\ 1964, NASA Special Publication, 50, 451




\bibitem[Chen(2011)]{2011LRSP....8....1C}
Chen, P.~F.\ 2011, Living Reviews in Solar Physics, 8, 1


\bibitem[Chen et al.(2007a)]{2007ApJ...665.1421C}
Chen, Y., Hu, Y.~Q.,\& Sun, S.~J.\ 2007a, \apj, 665, 1421


\bibitem[Chen et al.(2007b)]{2007AdSpR..40.1780C}
Chen, Y., Hu, Y.~Q.,\& Xia, L.~D.\ 2007b, Advances in Space
Research, 40, 1780


\bibitem[Cheng et al.(2014)]{2014ApJ...789L..35C}
Cheng, X., Ding, M.~D., Zhang, J., et al.\ 2014, \apjl, 789, L35


\bibitem[Cheng et al.(2013)]{2013ApJ...769L..25C}
Cheng, X., Zhang, J., Ding, M.~D., et al.\ 2013, \apjl, 769, L25


\bibitem[D{\'e}moulin\& Aulanier(2010)]{2010ApJ...718.1388D}
D{\'e}moulin, P., \&Aulanier, G.\ 2010, \apj, 718, 1388


\bibitem[Fan\& Gibson(2007)]{2007ApJ...668.1232F}
Fan, Y., \& Gibson, S.~E.\ 2007, \apj, 668, 1232


\bibitem[Forbes\& Isenberg(1991)]{1991ApJ...373..294F}
Forbes, T.~G., \&Isenberg, P.~A.\ 1991, \apj, 373, 294


\bibitem[Forbes\& Priest(1995)]{1995ApJ...446..377F}
Forbes, T.~G., \& Priest, E.~R.\ 1995, \apj, 446, 377






\bibitem[Hirayama(1974)]{1974SoPh...34..323H}
Hirayama, T.\ 1974, \solphys, 34, 323


\bibitem[Hu et al.(2003)]{2003JGRA..108.1072H}
Hu, Y.~Q., Li, G.~Q.,\& Xing, X.~Y.\ 2003, Journal of Geophysical
Research (Space Physics), 108, 1072


\bibitem[Isenberg et al.(1993)]{1993ApJ...417..368I}
Isenberg, P.~A., Forbes, T.~G., \& Demoulin, P.\ 1993, \apj, 417,
368


\bibitem[Kliem \& T\" or\" ok (2006)]{2006PhRvL..96y5002K}
Kliem, B., \& T\" or\" ok, T.\ 2006, Physical Review Letters, 96,
255002


\bibitem[Kopp\& Pneuman(1976)]{1976SoPh...50...85K}
Kopp, R.~A., \& Pneuman, G.~W.\ 1976, \solphys, 50, 85


\bibitem[Lemen et al.(2012)]{2012SoPh..275...17L}
Lemen, J.~R., Title, A.~M., Akin, D.~J., et al.\ 2012, \solphys,
275, 17

\bibitem[Lin (2002)]{Lin 2002}
Lin J.\ 2002, Chin. J. Astron. Astrophys., 2, 539

\bibitem[Lin \& Forbes (2000)]{Lin 2000}
Lin J., \& Forbes, T. G.\ 2000, \jgr, 105, 2375

\bibitem[Lin et al. (2003)]{Lin 2003}
Lin, J., Soon, W., Baliunas, S. L.\ 2003, New Astronomy Reviews,
47, 53

\bibitem[Lin et al.(2002)]{2002SoPh..210....3L}
Lin, R.~P., Dennis, B.~R., Hurford, G.~J., et al.\ 2002, \solphys,
210, 3


\bibitem[Mari\v ci\'c et al.(2007)]{2007SoPh..241...99M}
Mari\v ci\'c, D., Vr\v snak, B., Stanger, A.~L., et al.\ 2007,
\solphys, 241, 99


\bibitem[McKenzie\& Canfield(2008)]{2008A&A...481L..65M}
McKenzie, D.~E., \&Canfield, R.~C.\ 2008, \aap, 481, L65


\bibitem[Olmedo\& Zhang(2010)]{2010ApJ...718..433O}
Olmedo, O., \& Zhang, J.\ 2010, \apj, 718, 433


\bibitem[Priest\& Forbes(1990)]{1990SoPh..126..319P}
Priest, E.~R., \& Forbes, T.~G.\ 1990, \solphys, 126, 319

\bibitem[Priest\& Forbes(2000)]{Priest 2000}
Priest, E.~R., \& Forbes, T.~G.\ 2000, The Astronomy and
Astrophysics Review, 10, 313

\bibitem[Qiu et al.(2007)]{2007ApJ...659..758Q}
Qiu, J., Hu, Q., Howard, T.~A., \& Yurchyshyn, V.~B.\ 2007, \apj,
659, 758


\bibitem[Qiu et al.(2004)]{2004ApJ...604..900Q}
Qiu, J., Wang, H., Cheng, C.~Z., \& Gary, D.~E.\ 2004, \apj, 604,
900


\bibitem[Roussev et al.(2012)]{2012NatPh...8..845R}
Roussev, I.~I., Galsgaard, K., Downs, C., et al.\ 2012, Nature
Physics, 8, 845


\bibitem[Rust\& Kumar(1994)]{1994SoPh..155...69R}
Rust, D.~M., \& Kumar, A.\ 1994, \solphys, 155, 69





\bibitem[Song et al.(2015]{Song2015}
Song, H.~Q., Chen, Y., Zhang, J., et al.\ 2015, \apj, in press

\bibitem[Song et al.(2014a)]{2014ApJ...792L..40S}
Song, H.~Q., Zhang, J., Chen, Y., \& Cheng, X.\ 2014a, \apjl, 792,
L40


\bibitem[Song et al.(2014b)]{2014ApJ...784...48S}
Song, H.~Q., Zhang, J., Cheng, X., et al.\ 2014b, \apj, 784, 48


\bibitem[Sturrock(1966)]{Sturrock 1966}
Sturrock, P.~A.\ 1966, Natur, 211, 695




\bibitem[Titov\& D{\'e}moulin(1999)]{1999A&A...351..707T}
Titov, V.~S., \&D{\'e}moulin, P.\ 1999, \aap, 351, 707


\bibitem[van Tend\& Kuperus(1978)]{1978SoPh...59..115V}
van Tend, W., \& Kuperus, M.\ 1978, \solphys, 59, 115

\bibitem[Vr\v snak et al. (2005)]{Vrsnak 2005}
Vr\v snak, B., Sudar, D., \& Ru\v zdjak, D. 2005, \aap, 435, 1149

\bibitem[Wang\& Stenborg(2010)]{2010ApJ...719L.181W}
Wang, Y.-M., \& Stenborg, G.\ 2010, \apjl, 719, L181





\bibitem[Yashiro et al. (2002)]{Yashiro 2002}
Yashiro, S., Gopalswamy, N., Michalek, G., \& Howard, R. A. 2002,
BAAS, 34, 695


\bibitem[Zhang et al.(2001)]{2001ApJ...559..452Z}
Zhang, J., Dere, K.~P., Howard, R.~A., Kundu, M.~R., \& White,
S.~M.\ 2001, \apj, 559, 452


\bibitem[Zhang et al.(2012)]{2012NatCo...3E.747Z}
Zhang, J., Cheng, X., \& Ding, M.-D.\ 2012, Nature Communications,
3, 747








































































































\end{thebibliography}
\end{document}